\documentclass[a4paper,8pt]{article}
\usepackage[english]{babel}
\usepackage{graphicx}
\usepackage{amsmath}
\input{pcpdoc.sty}

\newcommand{\TITLE}{Advanced holographic nondestructive testing system for residual stress analysis}

\begin{document}

\begin{center}
{\LARGE\bf\TITLE}

\vspace{0.5cm}

{\bf Anatoli Kniazkov, Yuri Onischenko, George Dovgalenko, Gregory Salamo}

{\it Laser Laboratory, Physics Department, University of Arkansas}

{\it 226 Physics Bld, Fayetteville, AR, 72701, USA}

\vspace{0.5cm}

{\bf Tatiana Latychevskaia}

{\it St. Petersburg State Technical University}

{\it Politechnicheskaia 29, St. Petersburg, Russia, 195251}

\end{center}

\vspace{1cm}
%%%%%%%%%%%%%%%%%%%%%%%%%%%%%%%%%%%%%%%%%%%%%%%%%%%%%%%%%%%%%%%%%%%%%%%%%
%%%%%%%%%%%%%%%%%%%%%%%%%%%%%%%%%%%%%%%%%%%%%%%%%%%%%%%%%%%%%%%%%%%%%%%%%
%%%%%%%%%%%%%%%%%%%%%%%%%%%%%%%%%%%%%%%%%%%%%%%%%%%%%%%%%%%%%%%%%%%%%%%%%
\begin{center}
{\Large\bf{ABSTRACT}}
\end{center}
\vspace{1cm}

The design and operating of a portable holographic interferometer for residual stress analysis by creating a small scratch along with a new mathematical algorithm of calculations are discussed. Preliminary data of the stress investigations on aluminum and steel alloys have been obtained by the automatic processing of the interference pattern using a notebook computer. A phase-shift compensation technique in real-time reflection interferometry is used to measure the out-of-plane stress release surface displacement surrounding a small scratch (25~$\rm\mu$m depth and 0.5~mm width) in a plate with residual stress of around 50~MPa. Comparison between theoretical models for a rectangular and triangular shaped scratch with the experimental data are presented.

\vspace{0.5cm}

{\bf Keywords:} Displacement measurement, holographic interferometry, portable stress measurement

%%%%%%%%%%%%%%%%%%%%%%%%%%%%%%%%%%%%%%%%%%%%%%%%%%%%%%%%%%%%%%%%%%%%%%%%%
%%%%%%%%%%%%%%%%%%%%%%%%%%%%%%%%%%%%%%%%%%%%%%%%%%%%%%%%%%%%%%%%%%%%%%%%%
%%%%%%%%%%%%%%%%%%%%%%%%%%%%%%%%%%%%%%%%%%%%%%%%%%%%%%%%%%%%%%%%%%%%%%%%%
\vspace{1cm}
\begin{center}
{\Large\bf{1. INTRODUCTION}}
\end{center}
\vspace{1cm}

\begin{figure}[htbp]
  \begin{center}
 \includegraphics[width=10cm]{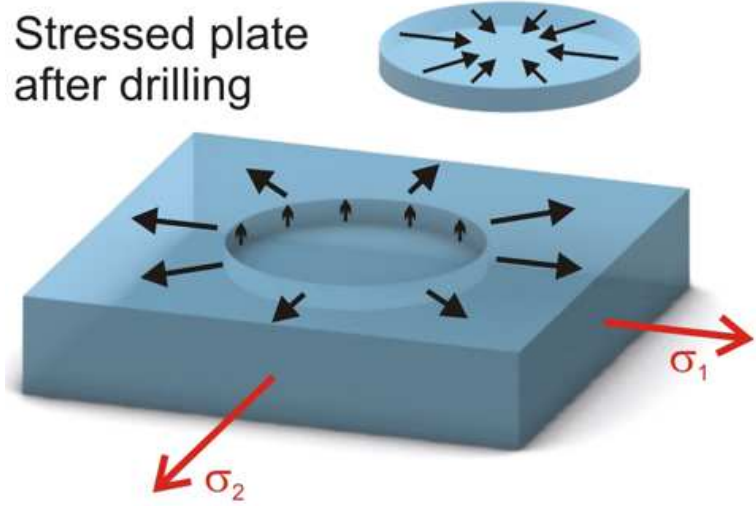}
    \parbox{0.8\textwidth}
    {\caption{Stress release displacement}
    \label{Fig:1}
      }
  \end{center}
\end{figure}

While the holographic technique sounds simple, it traditionally suffers from several serious problems. In particular, noise, vibration, air current, and temperature swings can prevent the capture of a hologram of the object. One approach is to use a short laser pulse to capture a hologram before movement can occur in such an environment. This approach, however, fails to allow a real-time comparison between a hologram of an object and the real image of the perturbed object. The difficulty is that although the short pulse hologram is successfully stored, movement due to the environment will take place before the object is altered and a comparison is made. Another problem which has traditionally plagued the holographic sensing is that techniques to perturb the structure have relied on drilling a hole in the structure (Fig. \ref{Fig:1}) in order to allow the internal stress to cause surface movement and, hence, change - hardly non-destructive!

%%%%%%%%%%%%%%%%%%%%%%%%%%%%%%%%%%%%%%%%%%%%%%%%%%%%%%%%%%%%%%
%%%%%%%%%%%%%%%%%%%%%%%%%%%%%%%%%%%%%%%%%%%%%%%%%%%%%%%%%%%%%%

Drilling a hole in a structure is very intimidating and usually unacceptable as a non-destructive test. In the approach taken in our work we demonstrate that only a small scratch is necessary for the measurement. In addition, we have forced the reference and object optical beams to move together in order to capture a hologram and compare with a real image of the scratched object. As a result, our approach has been successful even in poor environmental conditions.
%%%%%%%%%%%%%%%%%%%%%%%%%%%%%%%%%%%%%%%%%%%%%%%%%%%%%%%%%%%%%%%%%%%%%%%%%
%%%%%%%%%%%%%%%%%%%%%%%%%%%%%%%%%%%%%%%%%%%%%%%%%%%%%%%%%%%%%%%%%%%%%%%%%
%%%%%%%%%%%%%%%%%%%%%%%%%%%%%%%%%%%%%%%%%%%%%%%%%%%%%%%%%%%%%%%%%%%%%%%%%
\newpage

\begin{center}
{\Large\bf{2. DISCUSSION OF THE ALGORITHM}}
\end{center}
\vspace{1cm}

In this paper we present a theoretical analysis of the deformation created by a rectangular scratch on a surface as shown in Fig. 2.

%%%%%%%%%%%%%%%%%%%%%%%%%%%%%%%%%%%%%%%%%%%%%%%%%%%%%%%%%%%%%%
\begin{figure}[htbp]
  \begin{center}
 \includegraphics[width=10cm]{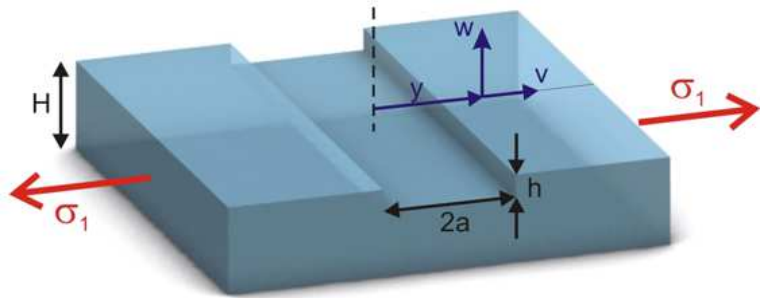}
    \parbox{0.8\textwidth}
    {\caption{Displacement components near the edge of the scratch produced in a stressed plate}
    \label{Fig:2}
      }
  \end{center}
\end{figure}
%%%%%%%%%%%%%%%%%%%%%%%%%%%%%%%%%%%%%%%%%%%%%%%%%%%%%%%%%%%%%%

We define three areas along the plate, as shown in Fig. 3 \cite{Kniazkov:1997}. In the first area (I) we have stressed the plate with a slot (the solution is well known), for the second area: (II) we have stressed the plate with a scratch and we will find the solution, and in the third area: (III) we have stressed the plate without a scratch (the solution is also well known). The solution for the radial component of the average values of the stresses, $\tilde{\sigma}_r$, and for out-of plane stress $\tilde{\tau}_z$ respectively, near the scratch produced in a finite plate under the state of stress $\sigma_1$ are expected in cylindrical coordinates as

\beq \tilde{\sigma}_r = \left( \frac{h+\Delta}{H} \right)\left[ -\frac{a^2}{r^2}n - \delta\left(-\frac{4a^2}{r^2} + \frac{3a^4}{r^4} \right)\cos{2\theta} \right] \nonumber\eeq

\beq\label{Eq:01} \tilde{\tau}_z = \left( \frac{h+\Delta}{H} \right)\left[  \delta\left(-\frac{3a^4}{r^4} + \frac{2a^2}{r^2} \right)\sin{2\theta} \right]  \eeq

where

\beq n = \frac{\sigma_1 + \sigma_2}{2}, \; \delta = \frac{\sigma_2 - \sigma_1}{2}, \nonumber \eeq

$h$ - the depth of a scratch, $H$ - the depth of the influence of a scratch releasing, $a$ - half of the scratch width, $r$ - current coordinate, $\Delta = (\Delta_2 - \Delta_1)/2$ - form-factor.

%%%%%%%%%%%%%%%%%%%%%%%%%%%%%%%%%%%%%%%%%%%%%%%%%%%%%%%%%%%%%%
\begin{figure}[htbp]
  \begin{center}
 \includegraphics[width=10cm]{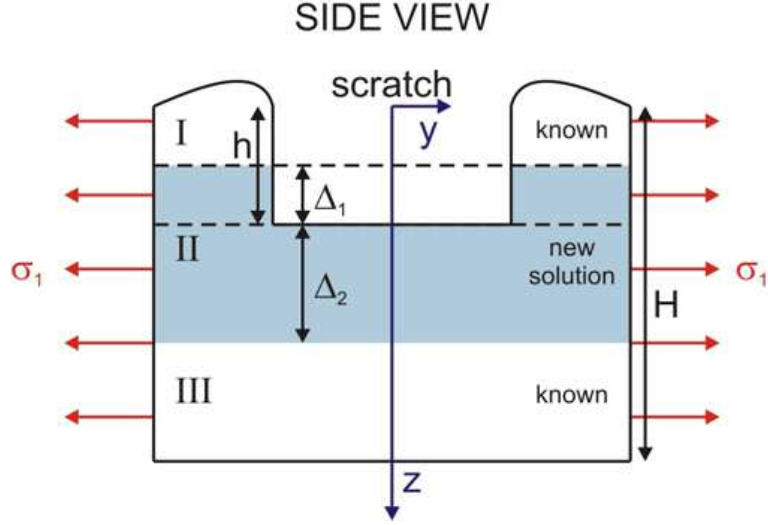}
    \parbox{0.8\textwidth}
    {\caption{Three areas configuration of stressed the plate model}
    \label{Fig:3}
      }
  \end{center}
\end{figure}
%%%%%%%%%%%%%%%%%%%%%%%%%%%%%%%%%%%%%%%%%%%%%%%%%%%%%%%%%%%%%%

The solution for the radial, tangential, and out-of-plane stress components near a scratch produced in a finite plate under biaxial stress can be expressed as a function of $\varphi$:

\beq\label{Eq:02} \sigma_x = \partdt{^2\varphi}{x^2};\qquad \sigma_y = \partdt{^2\varphi}{y^2};\qquad \tau_{xy} = - \partdt{^2\varphi}{x\ptdrv{y}} + qx; \eeq

where $\varphi$ satisfies the equation:

\beq\label{Eq:03} \partdt{^4\varphi}{y^4} + 2\partdt{^4\varphi}{x^2 {\ptdrv y^2}} + \partdt{^4\varphi}{x^4} = 0; \eeq

The solutions can be written as third degree of polynomial functions:

\beq\label{Eq:04} \varphi(x,y) = \sum_{\rm n=0,\; m=0}^{3} C_{\rm n,m} x^{\rm n} y^{\rm m}; \qquad {\rm n} + {\rm m} \leq 3; \qquad {\rm n}>0, \qquad {\rm m}>0; \eeq

In polar coordinates the equation for $\varphi$ becomes:

\beq\label{Eq:05} \left(\partdt{^2}{r^2} + \frac{1}{r} \partd{r} + \frac{1}{r^2} \partdt{^2}{\theta^2} \right)^2 \varphi = 0.\eeq

We will assume that a solution can be separated as:

\beq\label{Eq:06} \varphi(r,\theta) = f(r) \psi(\theta)\eeq

where

\beq\label{Eq:07} f(r) = r^{\rm k}, \qquad \psi(\theta) {\sim} \{ \cos({\rm m}\theta) +  \sin({\rm m}\theta) \} \eeq

In this case the general solution becomes:

\beq\label{Eq:08} \varphi(r,\theta) = \sum_{\rm m=0}^{\infty} \left( A_{\rm m}r^{\rm m} + B_{\rm m}r^{-\rm m} + C_{\rm m}r^{2 +\rm m} + D_{\rm m}r^{2 -\rm m}\right) \left( K_{\rm m}\cos({\rm m}\theta) + N_{\rm m} \sin({\rm m}\theta) \right) \eeq

%%%%%%%%%%%%%%%%%%%%%%%%%%%%%%%%%%%%%%%%%%%%%%%%%%%%%%%%%%%%%%%%%%%%%%%%%
%%%%%%%%%%%%%%%%%%%%%%%%%%%%%%%%%%%%%%%%%%%%%%%%%%%%%%%%%%%%%%%%%%%%%%%%%
%%%%%%%%%%%%%%%%%%%%%%%%%%%%%%%%%%%%%%%%%%%%%%%%%%%%%%%%%%%%%%%%%%%%%%%%%

\beq A_1 = A_1 K_1;\; B_1 = B_1 K_1;\; C_1 = C_1 K_1;\; D_1 = D_1 K_1;  \nonumber \eeq
\beq\label{Eq:09} A_2 = A_2 K_2;\; B_2 = B_2 K_2;\; C_2 = C_2 K_2;\; D_2 = D_2 K_2; \eeq
\beq A_3 = A_1 N_1;\; B_3 = B_1 N_1;\; C_3 = C_1 N_1;\; D_3 = D_1 N_1;  \nonumber \eeq
\beq A_4 = A_2 N_2;\; B_4 = B_2 N_2;\; C_4 = C_2 N_2;\; D_4 = D_2 N_2; \nonumber \eeq

It can be shown that:

\beq \varphi(r,\theta) = A_0 + (C_0 + A_2)x^2 + (C_0 - A_2)y^2 + A_1 x + A_3 y + C_1 x^3 + \nonumber\eeq
\beq\label{Eq:10} + C_3 y^3 + A_3 x y^2 + C_2 y x^2 + C_2 x^4 - C_2 y^4 + 2 A_4 y x + 2 C_4 y x^3 + 2 C_4 x y^3 + \eeq
\beq + \frac{1}{(x^2+y^2)} \left(B_1 x + B_3 y + (B_2 + D_2) x^2 + (B_2 - D_2) y^2 + 2 D_4 x y^2\right) + \nonumber\eeq
\beq + \frac{1}{(x^2+y^2)^2} \left(B_2 x^2 - B_2 y^2 + 2 B_4 x y\right). \nonumber\eeq

It follows that if we know the stress function $\varphi$, we can calculate the stress components using expression (\ref{Eq:02}):

\beq\label{Eq:11} \sigma_x = \partdt{^2\varphi}{x^2} = f_1 (A_{\rm i}, B_{\rm i}, C_{\rm i}, D_{\rm i}, x^{\rm i} y^{\rm i});\eeq

\beq\label{Eq:12} \sigma_y = \partdt{^2\varphi}{y^2} = f_2 (A_{\rm i}, B_{\rm i}, C_{\rm i}, D_{\rm i}, x^{\rm i} y^{\rm i});\eeq

\beq\label{Eq:13} \tau_{xy} = -\partdt{^2\varphi}{x\ptdrv{y}} + qx = f_3 (A_{\rm i}, B_{\rm i}, C_{\rm i}, D_{\rm i}, x^{\rm i} y^{\rm i});\eeq

where ${\rm i} = 0,1,2,3,4$; ${\rm j} = 0,1,2,3,4$;

We will find the solution in area (I) by taking into account the fact that the distribution of stresses must be equal for each plane ZY along direction X. For example, if we know the solution in the plane X$=0$ then it must be the same for each plane.

By calculating the stress using (\ref{Eq:11}), (\ref{Eq:12}), (\ref{Eq:13}), and taking $x = 0$, we get:

\beq\label{Eq:14} \sigma_x = 2 (C_0 - A_2) + 6C_3 y - 12 C_2 y^2 + 2\frac{B_3}{y^3} - 6 \frac{B_2}{y^4};\eeq

\beq\label{Eq:15} \sigma_y = 2 (C_0 + A_2) + 2C_5 y + 4 D_2 y^2 - 2\frac{B_3}{y^3} + 6 \frac{B_2}{y^4};\eeq

From $\sigma_2 \equiv 0$ it follows that:

\beq C_0 = A_2;\qquad  C_3 = C_2 = B_3 = B_2 = 0;\nonumber \eeq

Therefore we have:

\beq \sigma_x = 0;\nonumber\eeq

\beq\label{Eq:16} \sigma_y = 4 C_0 + 4\frac{D_2}{y^2} + 2 C_5 y;\eeq

The constants $C_0, D_2, C_5$ can also be found from the boundary conditions:

\beq \sigma_{y = \infty} = \sigma_1 = 4 C_0 + 4\frac{D_2}{y^2} + 2 C_5 y \; \Rightarrow  \;   C_0 = \frac{\sigma_1}{4};\; C_5 = 0;\nonumber\eeq

\beq\label{Eq:17} \sigma_{y = a} = \sigma_1 + 4\frac{D_2}{a^2}\; \Rightarrow  \; D_2 = - \sigma_1 \frac{a^2}{4};\eeq

%%%%%%%%%%%%%%%%%%%%%%%%%%%%%%%%%%%%%%%%%%%%%%%%%%%%%%%%%%%%%%%%%%%%%%%%%
%%%%%%%%%%%%%%%%%%%%%%%%%%%%%%%%%%%%%%%%%%%%%%%%%%%%%%%%%%%%%%%%%%%%%%%%%
%%%%%%%%%%%%%%%%%%%%%%%%%%%%%%%%%%%%%%%%%%%%%%%%%%%%%%%%%%%%%%%%%%%%%%%%%
\newpage

Thus the stresses are as follows:

\beq\label{Eq:18} \sigma_y = \sigma_1 \left(1 - \frac{a^2}{y^2}\right);\qquad \tau_{xy} = -2 \sigma_1 a^2 xy\left(\frac{x^2 - y^2}{x^2 + y^2}\right).\eeq

It is evident that for the area (III) we will have:

\beq \sigma_y = \sigma_1;\qquad \tau_{xy} = 0.\nonumber\eeq

Using the result from the method of quasi-generalized plane stress condition for a blind hole \cite{Kniazkov:1997}, we can obtain the solutions for a rectangular scratch:

\beq \sigma_y^{*} = \frac{1}{H}\left( \sigma_y^{\rm k} (h + \Delta) + \sigma_y^0 (H - h - \Delta) \right) = \nonumber\eeq

\beq\label{Eq:19}  = \frac{1}{H}\left( \sigma_1 \left(1 - \frac{a^2}{y^2}\right) (h + \Delta) + \sigma_1 (H - h - \Delta) \right) = \eeq

\beq  =  \sigma_1 - \sigma_1 \frac{(h + \Delta)}{H}\frac{a^2}{y^2}.\nonumber\eeq

The net result is:

\beq\label{Eq:20} \sigma_y^{\sim} = \sigma_y^{*} - \sigma_y^0 = \sigma_y^{*} - \sigma_1 = - \sigma_1 \frac{(h + \Delta)}{H}\frac{a^2}{y^2}.\eeq

For in plane and out of plane deformations we will have:

\beq \epsilon_y = \frac{\delta v}{\delta y} = \frac{1}{E} \left( \sigma_y - \nu (\sigma_z + \sigma_x) \right) = \frac{1}{E} \left( \sigma_y - \nu \sigma_z \right);\nonumber\eeq

\beq\label{Eq:21} \epsilon_x = \frac{\delta w}{\delta z} = \frac{1}{E} \left( \sigma_z - \nu (\sigma_y + \sigma_z) \right) = \frac{1}{E} \left( \sigma_z - \nu \sigma_y \right);\eeq

Therefore for in plane and out of plane displacements we get:

\beq\label{Eq:22} V = \int \epsilon_x {\rm d}y = \sigma_1 \frac{(h + \Delta)}{EH}\frac{a^2}{y};\eeq

\beq W = \int \epsilon_z {\rm d}z = \sigma_1 \nu h\frac{(h + \Delta)}{EH}\frac{a^2}{y^2};\nonumber\eeq

where $E$ is Young's modulus and $\nu$ is Poisson's ratio.
%%%%%%%%%%%%%%%%%%%%%%%%%%%%%%%%%%%%%%%%%%%%%%%%%%%%%%%%%%%%%%%%%%%%%%%%%
%%%%%%%%%%%%%%%%%%%%%%%%%%%%%%%%%%%%%%%%%%%%%%%%%%%%%%%%%%%%%%%%%%%%%%%%%
%%%%%%%%%%%%%%%%%%%%%%%%%%%%%%%%%%%%%%%%%%%%%%%%%%%%%%%%%%%%%%%%%%%%%%%%%
\newpage
\begin{center}
{\Large\bf{3. TRIANGULAR SCRATCH SOLUTION}}
\end{center}
\vspace{1cm}

\begin{figure}[htbp]
  \begin{center}
 \includegraphics[width=10cm]{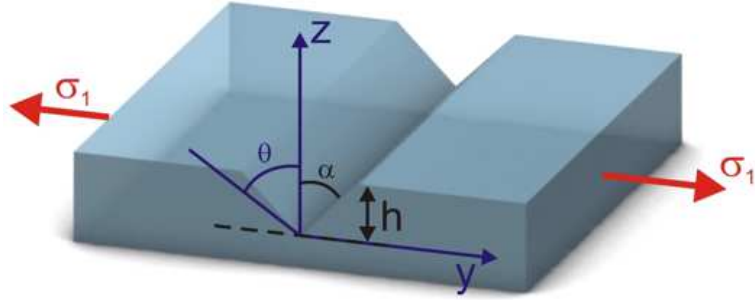}
    \parbox{0.8\textwidth}
    {\caption{Geometry of the triangular scratch}
    \label{Fig:4}
      }
  \end{center}
\end{figure}

For triangular scratch Fig. \ref{Fig:4} the stress function we will look for in terms \cite{Timoshenko:1970}:

\beq\label{Eq:23} \varphi(r,\theta) = r^{\lambda + 1} f(\theta)\eeq

\beq f(\theta) = C_1 \sin{((\lambda + 1)\theta)} + C_2 \cos{((\lambda + 1)\theta)} + C_3 \sin{((\lambda - 1)\theta)} + C_4 \cos{((\lambda - 1)\theta)}\nonumber\eeq

%%%%%%%%%%%%%%%%%%%%%%%%%%%%%%%%%%%%%%%%%%%%%%%%%%%%%%%%%%%%%%
%%%%%%%%%%%%%%%%%%%%%%%%%%%%%%%%%%%%%%%%%%%%%%%%%%%%%%%%%%%%%%

The stress and displacement components (ignoring rigid-body terms) are given by \cite{Timoshenko:1970}:

\beq \sigma_r = r^{\lambda -1} \left( f'(\theta) + (\lambda + 1) f(\theta)\right); \nonumber\eeq

\beq\label{Eq:24} \sigma_{\theta} = r^{\lambda -1} \lambda (\lambda + 1) f(\theta); \eeq

\beq \tau_{r\theta} = - r^{\lambda -1} \lambda  f'(\theta); \nonumber\eeq

\beq 2GV = r^{\lambda} \left( - f'(\theta) + \left(\frac{\lambda - 1}{1 + \nu}\right) g(\theta)\right); \nonumber\eeq

\beq 2GW = r^{\lambda} \left( -(\lambda + 1) f(\theta) + \left(\frac{1}{1 + \nu}\right) g'(\theta)\right); \nonumber\eeq

where

\beq g(\theta) = \frac{4}{\lambda - 1}\left( C_3 \cos{((\lambda - 1)\theta)} +  C_4 \sin{((\lambda - 1)\theta)} \right); \qquad G = \frac{E}{2(1 + \nu)}.\nonumber\eeq

Let's consider the wedge-shaped region bounded by radii $\theta = \pm\alpha$, which are to be free of load, so that $\sigma_{\theta} = 0$. This means $f(\alpha) = f(-\alpha) = 0$. Taking into account the fact that $f(\theta)$ is a symmetric function of $\theta$ we get $f'(\alpha) = f'(-\alpha)$ and $C_1 = C_3 = 0$, $\lambda = 1$.

\vspace{0.5cm}

Then stress function (\ref{Eq:24}) becomes

\beq\label{Eq:25} \phi (r, \theta) = r^3\left( C_2\cos{(2\theta)} + C_4\right). \eeq

Solving the system (\ref{Eq:25}) as we did before we get the stress components:

\beq \sigma_r = -\frac{\sigma_1}{2} \left( \cos{(2\theta)} + \cos{(2\alpha)}\right);\nonumber \eeq

\beq\label{Eq:26} \sigma_{\theta} = \frac{\sigma_1}{2} \left( \cos{(2\theta)} - \cos{(2\alpha)}\right); \eeq

\beq \tau_{r\theta} = \frac{\sigma_1}{2} \sin{(2\theta)};\nonumber \eeq

and for in plane and out of plane stress components we will find:

\beq \sigma_y = \frac{\sigma_1}{2} \left( 1 - \cos{(2\alpha)}\right);\nonumber \eeq

\beq\label{Eq:27} \sigma_z = - \frac{\sigma_1}{2} \left( 1 + \cos{(2\alpha)}\right). \eeq

\vspace{0.5cm}

Therefore, for in plane and out of plane displacements on the border of a scratch we get:

\beq V = \frac{\sigma_1 h \tan{\alpha}}{2E} \left( (1 + \nu) + (\nu - 1) \cos(2\alpha)\right); \nonumber\eeq

\beq\label{Eq:28} W = \frac{\sigma_1 h}{2E} \left( - (1 + \nu) + (\nu - 1) \cos(2\alpha)\right).\eeq

%%%%%%%%%%%%%%%%%%%%%%%%%%%%%%%%%%%%%%%%%%%%%%%%%%%%%%%%%%%%%%%%%%%%%%%%%
%%%%%%%%%%%%%%%%%%%%%%%%%%%%%%%%%%%%%%%%%%%%%%%%%%%%%%%%%%%%%%%%%%%%%%%%%
%%%%%%%%%%%%%%%%%%%%%%%%%%%%%%%%%%%%%%%%%%%%%%%%%%%%%%%%%%%%%%%%%%%%%%%%%
\newpage
\begin{center}
{\Large\bf{4. MEASUREMENT SETUP}}
\end{center}
\vspace{1cm}

We used a holographic interferometer to provide the means to detect surface displacements produced by the introduction of a scratch on the surface of the material under stress. In this technique a hologram is made before the surface is scratched and compared with the real-time image after a scratch is made on the surface. The reconstructed wave fronts from the original image of the surface and from the deformed image of the surface after the introduction of the scratch, will add together with a phase shift $\varphi (x,y)$ between caused by the displacement of the surface after being scratched. As a result of the phase shift interference fringes will be observed by viewing both images simultaneously.

The optical setup used in the tests is shown in Fig. 5. Before scratching a surface of the test object, a hologram was made of a given location on the test object. After a scratch is made we obtain using a CCD-camera the resulting interference fringe pattern that is immediately viewed in real time on a video monitor of a PC-computer. Variable phase shifts are introduced into the reference arm of the He-Ne laser after collimation using a piezoelectric transducer that is controlled by PC-computer to measure a small phase shift (producing less than one fringe) resulting from the scratch. These interference patterns of the test surface area entered in the PC-computer where they are processed and the principal stress of the test object are determined automatically.

%%%%%%%%%%%%%%%%%%%%%%%%%%%%%%%%%%%%%%%%%%%%%%%%%%%%%%%%%%%%%%
\begin{figure}[htbp]
  \begin{center}
 \includegraphics[width=12cm]{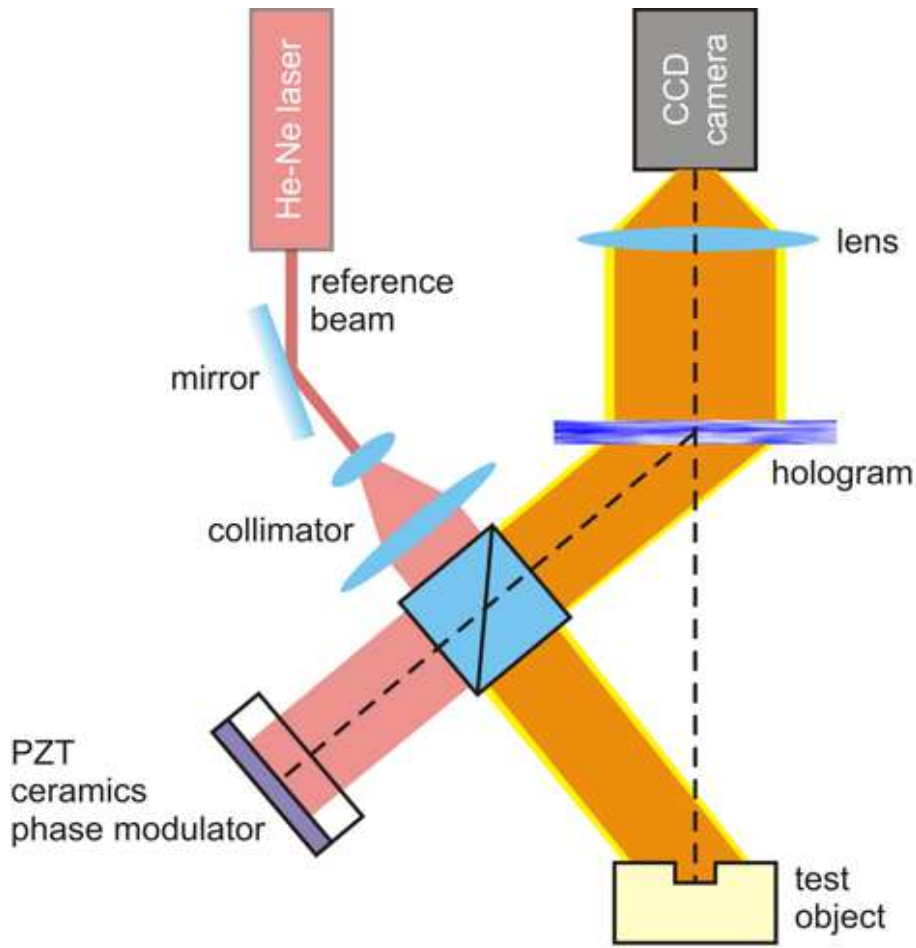}
    \parbox{0.8\textwidth}
    {\caption{Schematic of the apparatus}
    \label{Fig:5}
      }
  \end{center}
\end{figure}
%%%%%%%%%%%%%%%%%%%%%%%%%%%%%%%%%%%%%%%%%%%%%%%%%%%%%%%%%%%%%%

%%%%%%%%%%%%%%%%%%%%%%%%%%%%%%%%%%%%%%%%%%%%%%%%%%%%%%%%%%%%%%%%%%%%%%%%%
%%%%%%%%%%%%%%%%%%%%%%%%%%%%%%%%%%%%%%%%%%%%%%%%%%%%%%%%%%%%%%%%%%%%%%%%%
%%%%%%%%%%%%%%%%%%%%%%%%%%%%%%%%%%%%%%%%%%%%%%%%%%%%%%%%%%%%%%%%%%%%%%%%%
\newpage
\begin{center}
{\Large\bf{5. EXPERIMENTAL RESULTS AND CONCLUSIONS}}
\end{center}
\vspace{1cm}

Equations (\ref{Eq:22}) and (\ref{Eq:28}) give the prediction for $V$ and $W$ displacements resulting from scratching s stressed surface with a rectangular or triangular shaped scratch respectively. Both the rectangular and triangular results are similar.

\vspace{0.5cm}

The predictions are that $W$ will have a linear to quadratic dependence on the depth while being independent of the width. Fig. 6 shows that this is in fact exactly what we have observed experimentally. The three curves for a rectangular scratch are nearly (within experimental error) identical even though the width was varied from 2 to 4 mm and have a greater than linear dependence on depth. The fourth curve for a triangular scratch is also similar to the three curves for the rectangular scratch although a more slowly varying function of depth as predicted (this is also summarized in Fig. 8).

%%%%%%%%%%%%%%%%%%%%%%%%%%%%%%%%%%%%%%%%%%%%%%%%%%%%%%%%%%%%%%
\begin{figure}[htbp]
  \begin{center}
 \includegraphics[width=12cm]{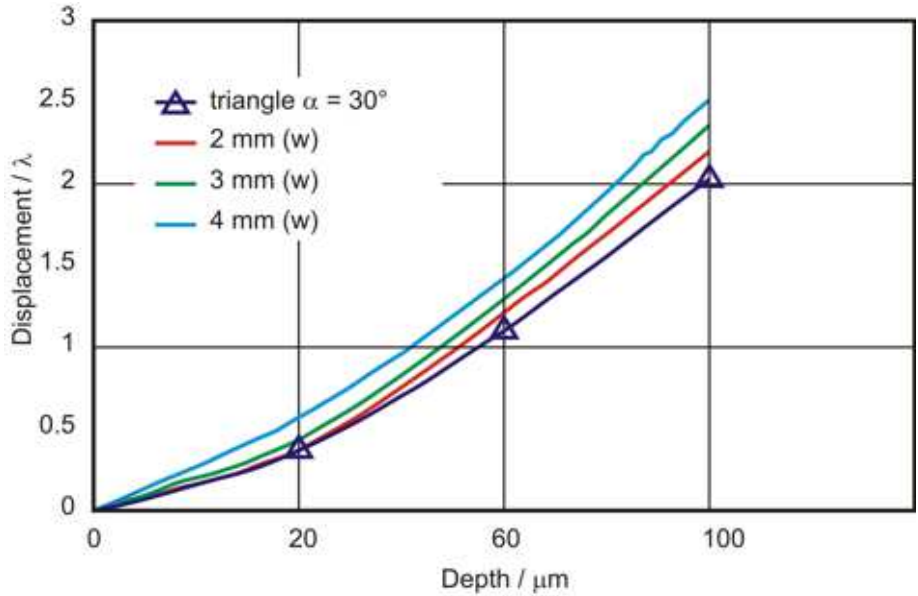}
    \parbox{0.8\textwidth}
    {\caption{The displacement ($W$) of the edge of the rectangular scratch versus it's depth and width}
    \label{Fig:6}
      }
  \end{center}
\end{figure}
%%%%%%%%%%%%%%%%%%%%%%%%%%%%%%%%%%%%%%%%%%%%%%%%%%%%%%%%%%%%%%
%%%%%%%%%%%%%%%%%%%%%%%%%%%%%%%%%%%%%%%%%%%%%%%%%%%%%%%%%%%%%%
\begin{figure}[htbp]
  \begin{center}
 \includegraphics[width=12cm]{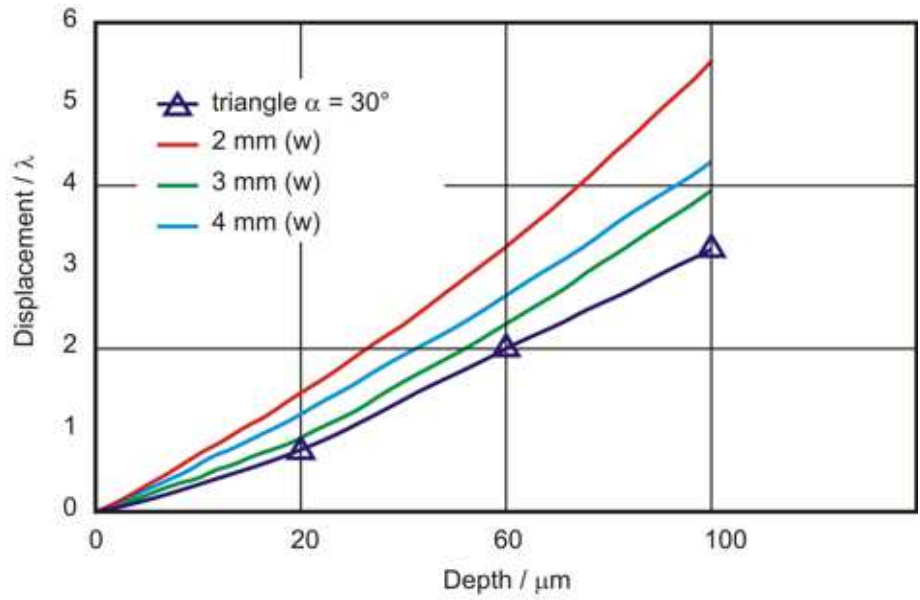}
    \parbox{0.8\textwidth}
    {\caption{The displacement ($V$) of the edge of the rectangular scratch, in the direction in the plane of the surface, versus the depth and width of the scratch}
    \label{Fig:7}
      }
  \end{center}
\end{figure}
%%%%%%%%%%%%%%%%%%%%%%%%%%%%%%%%%%%%%%%%%%%%%%%%%%%%%%%%%%%%%%
%%%%%%%%%%%%%%%%%%%%%%%%%%%%%%%%%%%%%%%%%%%%%%%%%%%%%%%%%%%%%%
\begin{figure}[htbp]
  \begin{center}
 \includegraphics[width=10cm]{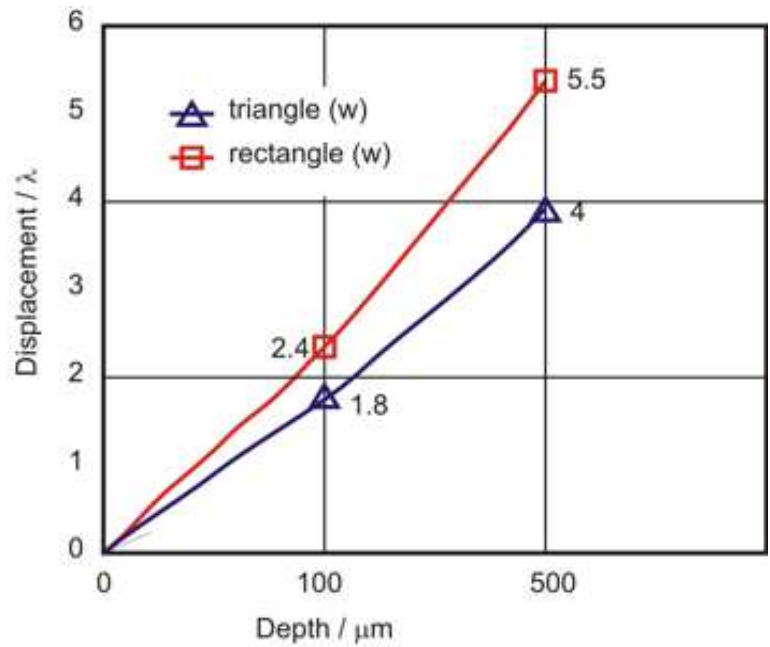}
    \parbox{0.8\textwidth}
    {\caption{Influence of the scratch shapes and depth}
    \label{Fig:8}
      }
  \end{center}
\end{figure}
%%%%%%%%%%%%%%%%%%%%%%%%%%%%%%%%%%%%%%%%%%%%%%%%%%%%%%%%%%%%%%

The predictions for $V$ are somewhat more complicated. For the rectangular scratch the predictions is a linear dependence on width. This is indeed the case as we have observed a near linear dependence on width shown in Fig. 7. The fourth curve, for a triangular scratch, is also in agreement with theory showing the predicted linear dependence on depth.

Figure 9 shows a picture of the holographic stress measurement device used for this study.

%%%%%%%%%%%%%%%%%%%%%%%%%%%%%%%%%%%%%%%%%%%%%%%%%%%%%%%%%%%%%%
\begin{figure}[htbp]
  \begin{center}
 \includegraphics[width=10cm]{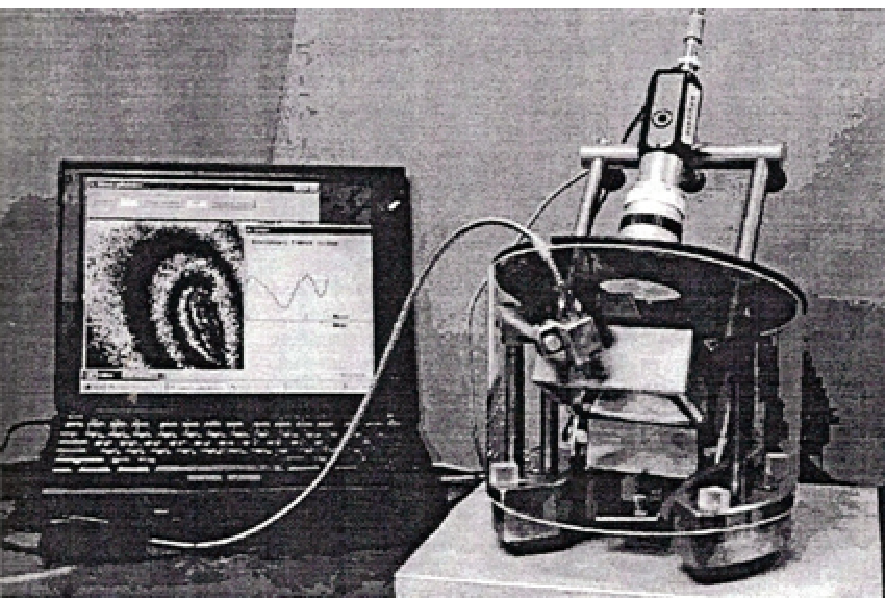}
    \parbox{0.8\textwidth}
    {\caption{Photograph of the portable holographic instrument for measurement of residual stress}
    \label{Fig:9}
      }
  \end{center}
\end{figure}
%%%%%%%%%%%%%%%%%%%%%%%%%%%%%%%%%%%%%%%%%%%%%%%%%%%%%%%%%%%%%%

%%%%%%%%%%%%%%%%%%%%%%%%%%%%%%%%%%%%%%%%%%%%%%%%%%%%%%%%%%%%%%%%%%%%%%%%%
%%%%%%%%%%%%%%%%%%%%%%%%%%%%%%%%%%%%%%%%%%%%%%%%%%%%%%%%%%%%%%%%%%%%%%%%%
%%%%%%%%%%%%%%%%%%%%%%%%%%%%%%%%%%%%%%%%%%%%%%%%%%%%%%%%%%%%%%%%%%%%%%%%%
\bibliographystyle{osa}
\bibliography{References_Holography}

\end{document}